\documentclass[journal]{IEEEtran}

\ifCLASSOPTIONcompsoc
  \usepackage[nocompress]{cite}
\else
  \usepackage{cite}
\fi
\ifCLASSINFOpdf
\else
\fi

\usepackage{graphicx}
\usepackage{amsmath,amssymb,amsfonts}
\usepackage{subcaption}
\usepackage{tabularx}
\usepackage{textcomp}
\usepackage{xcolor}
\usepackage{booktabs} 
\usepackage{balance}
\usepackage{algpseudocode}
\usepackage{color}
\usepackage{xcolor}
\usepackage{pifont}
\usepackage{soul}
\usepackage{url}
\usepackage{algorithm}
\usepackage{algorithmicx}
\usepackage{xurl}
\usepackage{hhline}
\usepackage{multirow}
\usepackage{makecell}
\usepackage{caption}
\usepackage{fancybox}
\usepackage{tikz}
\usepackage[flushleft]{threeparttable}
\usepackage{framed}
\usepackage{multicol}
\usepackage{rotating} 
\usepackage{lipsum}
\usepackage{array}
\usepackage{tikz}
\usepackage{colortbl}
\usepackage{amssymb}
\usepackage{pifont}
\newcommand{\xmark}{\ding{55}}%
\usepackage{wasysym}
\DeclareGraphicsExtensions{.eps,.ps,.jpg,.bmp,.pdf}
\graphicspath{{figure/}}

\newcommand*\halfcirc[1][1ex]{%
  \begin{tikzpicture}
  \draw[fill] (0,0)-- (90:#1) arc (90:270:#1) -- cycle ;
  \draw (0,0) circle (#1);
  \end{tikzpicture}}

\newcommand\mydots{\ifmmode\ldots\else\makebox[1em][c]{.\hfil.\hfil.}\fi}

\usepackage{amsthm}

\usepackage{float}
\usepackage{hyperref}
\hypersetup{colorlinks,linkcolor={blue},citecolor={blue},urlcolor={red}} 

\usepackage{cleveref}

\crefformat{section}{\S#2#1#3} 
\crefformat{subsection}{\S#2#1#3}
\crefformat{subsubsection}{\S#2#1#3}

\begin{document}

\title{Security and Privacy of 6G Federated Learning-enabled Dynamic Spectrum Sharing}

\author{
    \IEEEauthorblockN{Viet Vo\IEEEauthorrefmark{1}\IEEEauthorrefmark{2}, Thusitha Dayaratne\IEEEauthorrefmark{2},  
    Blake Haydon\IEEEauthorrefmark{2}, 
    Xingliang Yuan\IEEEauthorrefmark{2}\IEEEauthorrefmark{4},  
    Shangqi Lai\IEEEauthorrefmark{3}, 
    Sharif Abuadbba\IEEEauthorrefmark{3}, 
    Hajime Suzuki\IEEEauthorrefmark{3}, 
    Carsten Rudolph\IEEEauthorrefmark{2}
    }\\
        \IEEEauthorblockA{\IEEEauthorrefmark{1}Swinburne University of Technology, Australia
     }\\
    \IEEEauthorblockA{\IEEEauthorrefmark{2}Monash University, Australia
   }\\
    \IEEEauthorblockA{\IEEEauthorrefmark{3}Data61, CSIRO, Australia
     }\\

    \IEEEauthorblockA{\IEEEauthorrefmark{4}University of Melbourne, Australia
     }\\     

}

\markboth{Submitted to IEEE Networks}%
{Shell \MakeLowercase{\textit{et al.}}: A Sample Article Using IEEEtran.cls for IEEE Journals}

\maketitle

\begin{abstract}

Spectrum sharing is increasingly vital in 6G wireless communication, facilitating dynamic access to unused spectrum holes.
Recently, there has been a significant shift towards employing machine learning (ML) techniques for sensing spectrum holes. 
In this context, federated learning (FL)-enabled spectrum sensing technology has garnered wide attention, allowing for the construction of an aggregated ML model without disclosing the private spectrum sensing information of wireless user devices. 
However, the integrity of collaborative training and the privacy of spectrum information from local users have remained largely unexplored. 
This article first examines the latest developments in FL-enabled spectrum sharing for prospective 6G scenarios. 
It then identifies practical attack vectors in 6G to illustrate potential AI-powered security and privacy threats in these contexts. 
Finally, the study outlines future directions, including practical defense challenges and guidelines.

%
%

\end{abstract}


%

\section{Introduction}

Emerging from the successful global deployment of fifth-generation wireless communication, the upper mid-band spectrum airwaves ($7.1- 24$ GHz) continue to hold a pivotal role for the forthcoming generation, slated for deployment in the 2030s\cite{Americas23}. 
This spectrum, often referred to as the "Goldlock" band, strikes a delicate balance between the coverage, capacity and data transmission rates. For instance, LTE/5G radio cells boast speeds ranging from 2 to 20 Gbps within a 2-mile radius area.
This distinctive advantage renders the airwaves highly suitable for next-generation applications necessitating real-time responsiveness. These include enhanced mobile broadband, augmented reality, virtual reality, IoT devices, mission-critical services for autonomous vehicles, remote surgery, and factory automation.

\begin{figure}[!t]
\centering
\includegraphics[height=4.5cm]{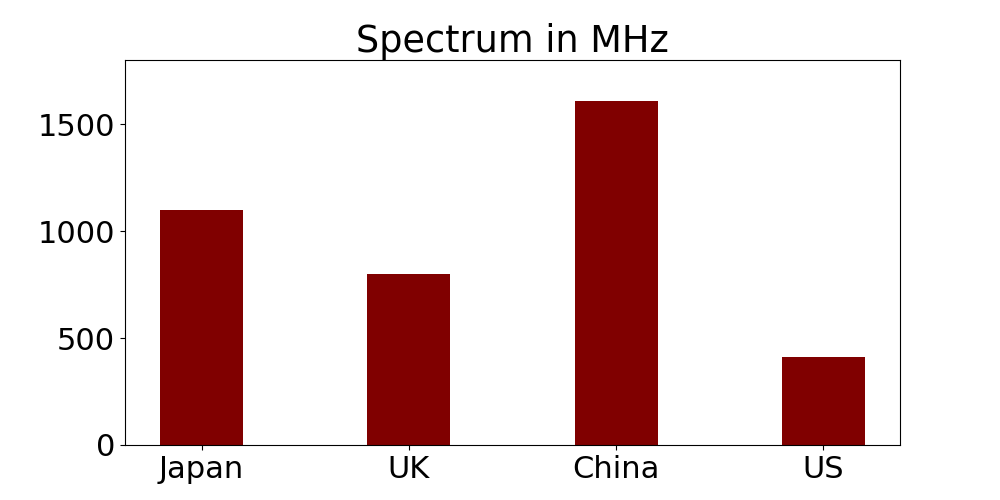}
\caption{Current and future mid-band spectrum allocation\cite{Americas23}.}
\label{fig:statistic}
\end{figure}

Given the critical role of upper mid-band airwaves, they are deemed a significant national resource. 
However, this valuable resource is finite, and the emergence of spectrum gaps among nations for licensed mid-band allocations hinders global technological advancement toward a connected future. 
Figure {\color{red}\ref{fig:statistic}} shows the mid-band spectrum shortage in certain countries including US and UK.
Consequently, the importance of spectrum sharing becomes evident in maximizing spectrum availability usage in the 6G era.
To this end, the IEEE Dynamic Spectrum Access Networks Committee (DySPAN-SC) plays a pivotal role in establishing spectrum management standards (IEEE P1900). 

A tangible example of spectrum sharing is the common 3.5 GHz band, shared by both U.S. Navy radar systems and commercial users under the Citizens Broadband Radio Service. 
To enable such spectrum sharing, a sensing network detects radar signals to reconfigure 5G communications of commercial users, preventing interference with radar signals.
In addition, leading telecommunication companies are extensively exploring spectrum-sharing models to have global impacts. 
This includes spectrum sharing between low-earth orbit (LEO) satellites and terrestrial networks or among industrial sensors and robotic systems/video surveillance within industrial facilities.
Therefore, promoting and fostering research and development activities for fine-grained and secure dynamic spectrum-sharing protocols are highlighted in the U.S. national spectrum strategy roadmap\cite{NTIA24}.

Traditional spectrum sensing technology detects unused spectrum frequency bands by examining the radio-environment map of signal energy in a coverage area.
However, this approach is unreliable due to the adverse propagation effects of wireless channels.
Furthermore, it does not meet the demand for dynamic spectrum access by wireless devices due to the lengthy examination process.

\begin{figure*}[!t]
\centering
\includegraphics[height=5.3cm]{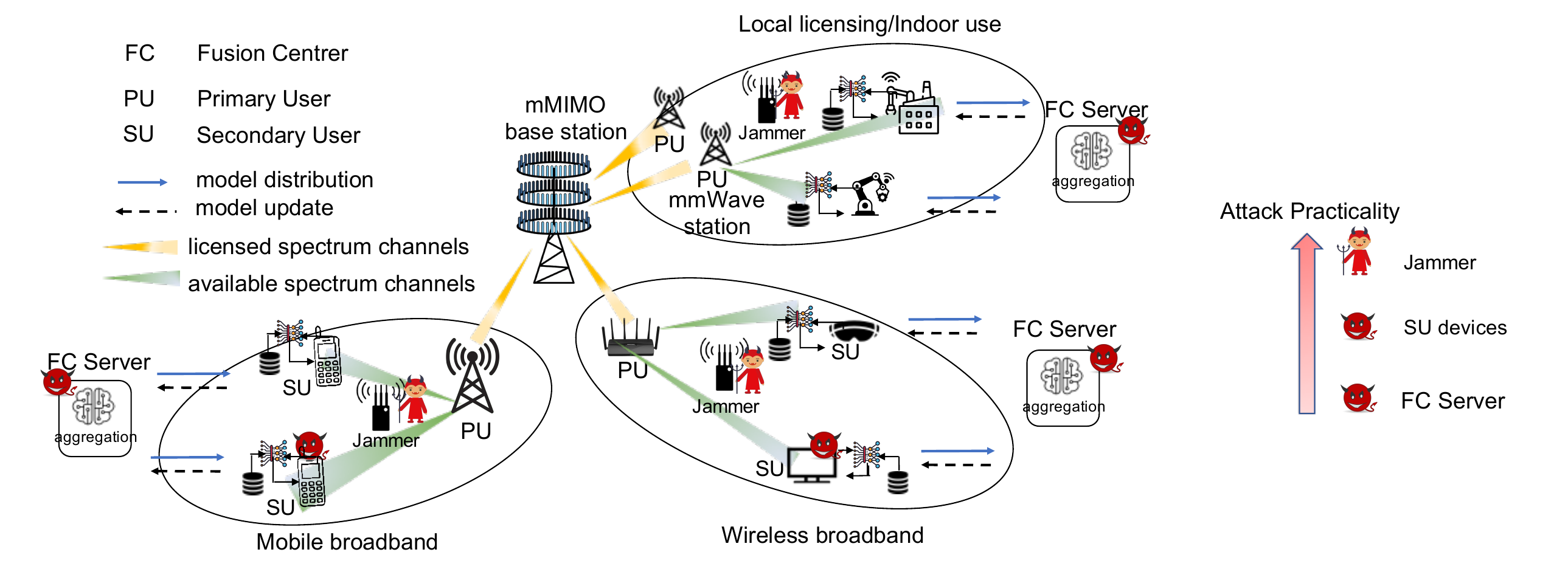} 
\caption{Threat landscape in FL-enabled dynamic spectrum sensing}
\label{fig:landscape}
\end{figure*} 

A modern spectrum sensing approach involves real-time measurement of radio transmission activities to support dynamic spectrum access.
Specifically, this sensing technique captures the time and resource allocation patterns of wireless devices in 5G systems.
This approach can be enhanced by employing machine learning (ML) algorithms to recognize resource allocation patterns, enabling the trained model to predict future spectrum occupation.
However, a centralized ML solution for training the model necessitates a large, localized spectrum sensing dataset gathered from numerous source devices, which may be impractical to acquire in real-time.
Furthermore, the reliability of the dataset depends on the quality of the wireless channels, including propagation effects.
Additionally, there are security and privacy concerns regarding sharing raw spectrum sensing information with the server for training such a centralized model.
For instance, the server can infer the geographic locations of source devices given the received signal strength via triangulation solution and infer the time-frequency traffic patterns of the devices\cite{Shi22,Wasilewska23}.

To address the privacy of source devices, federated learning (FL)\cite{McMahan2016}, a collaborative learning approach, has come across to enable the training of a global ML model without requiring the devices to reveal local spectrum sensing data.
In the FL paradigm, individual devices locally train an ML model using the local dataset so that the model's parameters are revealed to the server for a federated aggregation.
Adapting FL to the spectrum sensing systems do not only ensure the privacy of the local datasets but also provides communication efficiency and faster training latency compared to the centralized ML solution\cite{Wasilewska23}.
Given such promising benefits, there have been many studies that recently adapted FL to perform large-scale 6G spectrum sensing systems as listed in Table {\color{red}\ref{tab:attack_position}}.


Despite the extensive development of FL-based spectrum-sharing, the security and privacy of such collaborative training remain largely unexplored.
Recent FL studies in computer vision and language processing have raised alarm bells, revealing vulnerabilities to Byzantine poisoning attacks\cite{Minghong20} and privacy inference attacks \cite{Chaudhari23}. 
Specifically, while the former compromises the aggregation integrity of collaborative training, the latter targets the privacy of FL participants.
When extending FL to spectrum-sharing systems, the attack surface expands significantly, especially considering that 6G can connect up to $100$ million wireless devices per square kilometer according to ITU-R M. 2160.
The threat landscape in 6G is broader, encompassing not only compromised wireless devices participating in collaborative learning but also external adversarial devices, such as wireless jammers, interfering with communication between benign devices and the FL aggregation server.

The contribution of this article is a forward-looking exploration of the feasibility of migrating existing FL poisoning attacks and privacy inference attacks to 6G spectrum sharing.
In this context, the article addresses not only colluding Byzantine wireless devices but also disruptive wireless technologies.
Furthermore, this vision will highlight defense challenges and offer research recommendations aimed at fostering trustworthy and secure FL for spectrum sharing technologies.

\section{FL threat landscape in the mid-band spectrum}



\noindent \textbf{Cooperative Spectrum Sensing Adaptation}: Detecting available spectrum holes to maximize spectrum usage is not a new topic exclusive to 6G. 
Instead, it has a long history dating back to 5G and 5G Advanced deployments through cooperative spectrum sensing (CSS) design, which entails a centralized system comprising a primary user (PU), a fusion center (also known as an FC server), and a group of secondary users (SUs). 
At a high level, the SUs sense the frequently used channels and share that information with the FC, enabling it to determine and schedule the spectrum occupancy in a coverage area. 
Since the introduction of FL\cite{McMahan2016}, its adaptation to the CSS design has become more practical, not only because it protects the privacy of SUs' local training but also due to design compatibility. 
Specifically, the FC server can execute federated aggregation at every FL training round to generate a new global ML spectrum prediction model. 
%

 \begin{figure*}
    \centering
    \begin{subfigure}[t]{0.5\textwidth}
        \centering
        \includegraphics[height=4.9cm]{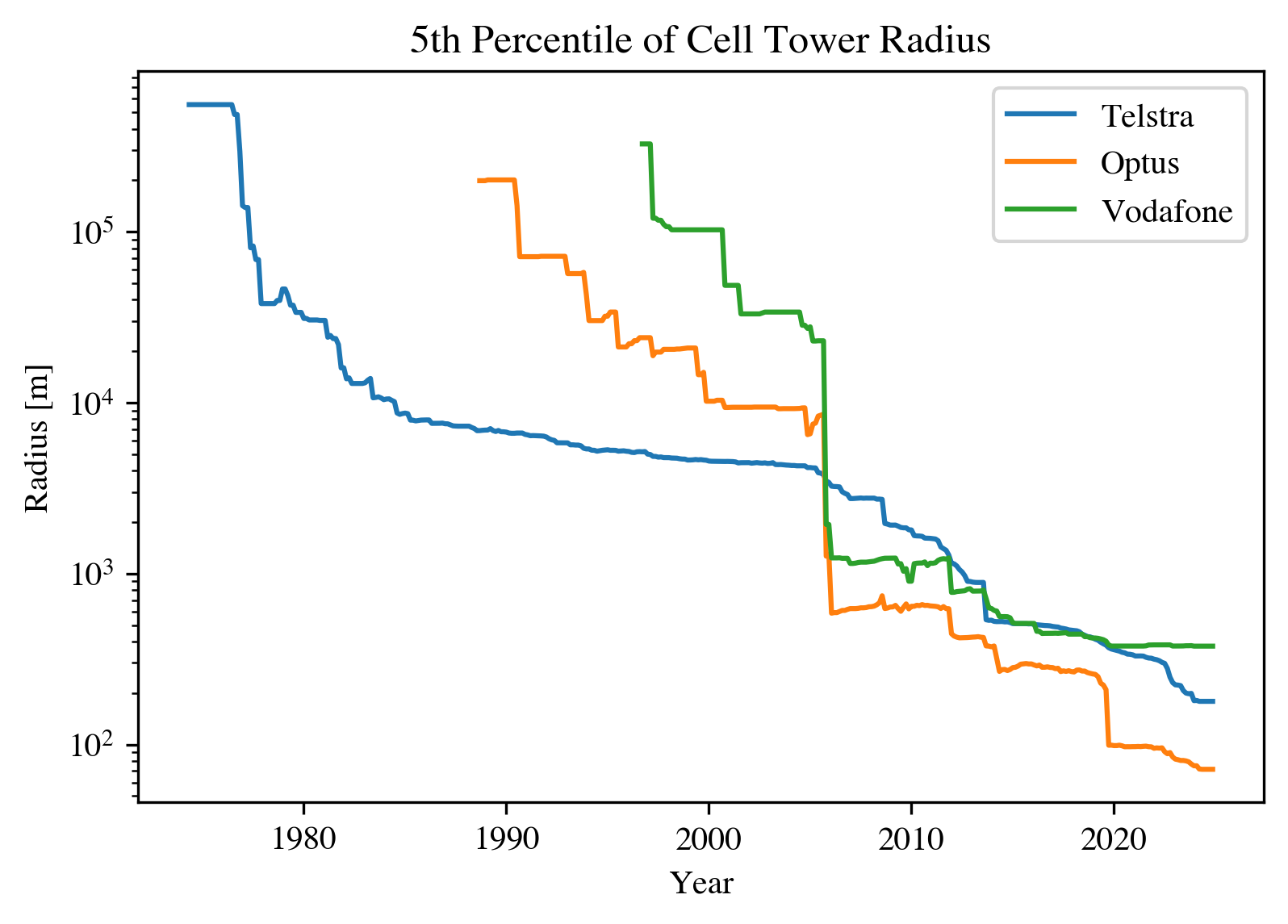}
        \caption{The reducing coverage radius of cell towers registered at Australian Communication and Media Authority (ACMA) from $1970$ to $2023$}
    \end{subfigure}%
    ~ 
    \begin{subfigure}[t]{0.5\textwidth}
        \centering
        \includegraphics[height=4.9cm]{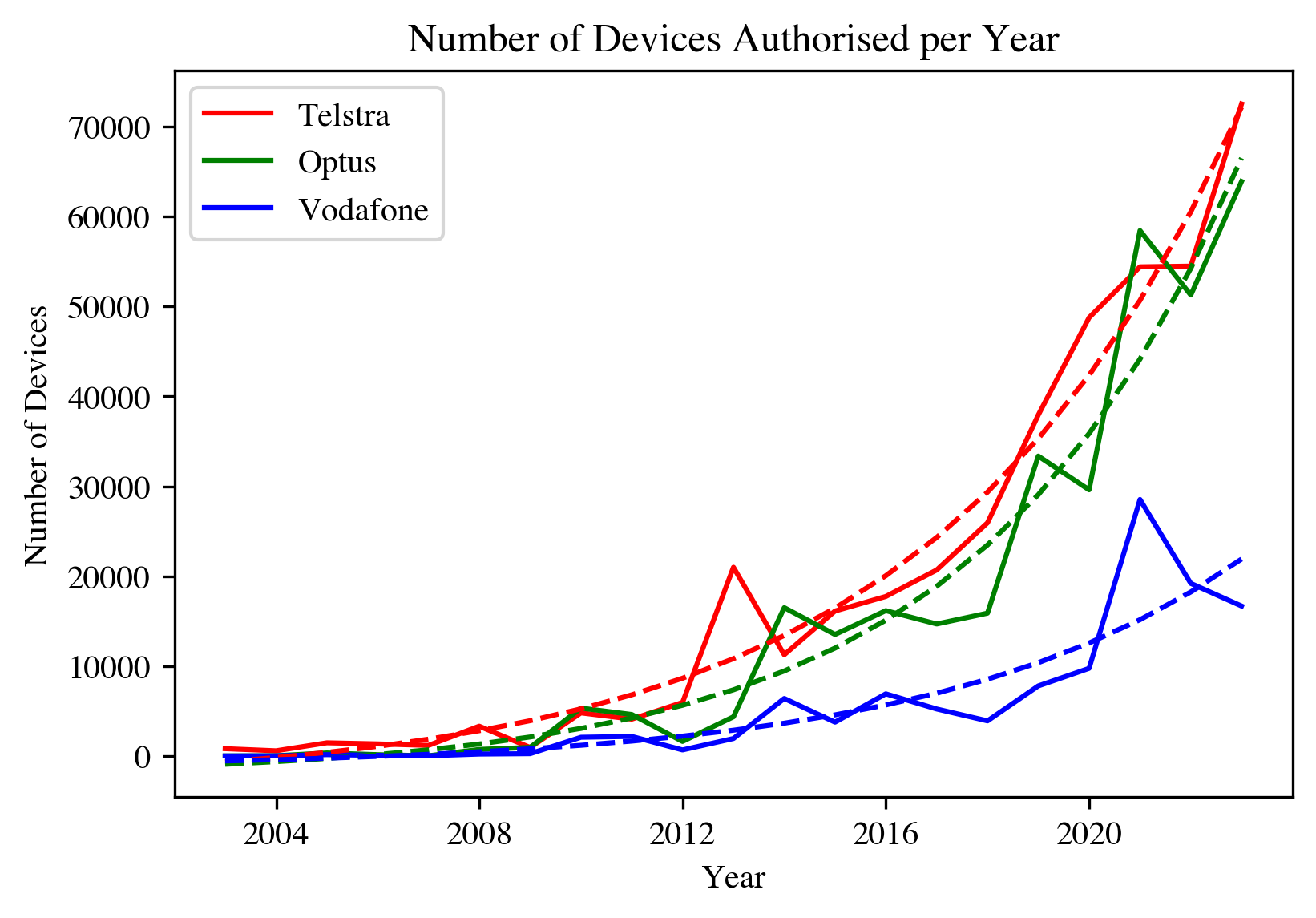}
        \caption{The increasing number of authorized devices from $2003$ to $2023$}
        \label{fig:devices_trends}
    \end{subfigure}
    \caption{Cell and devices trending through the lens of leading Australian telecom providers}
    \label{fig:cell_devices_trends}
\end{figure*}

\noindent \textbf{Real-world 6G spectrum sharing applications}: The deployment of the upper mid-band spectrum, ranging from $7.1$ to $24$ GHz, for 6G is now more practical and feasible than ever before. 
This is primarily due to the compatibility of these bands with the sub-6GHz frequencies used in existing 5G and 5G Advanced deployments. 
Moreover, these high-frequency bands have smaller wavelengths, enabling the integration of advanced antenna elements. 
For example, the $15$ to $24$ GHz band, with wavelengths between $1.25$ and $1.99$ cm, can utilize $0.13\mu$m Complementary metal–oxide–semiconductor (CMOS) 6-bit active digital phase shift technology. 
Therefore, these spectrum bands have the potential for terrestrial use cases that require high-bandwidth coverage, such as mobile broadband in urban areas and indoor and outdoor-to-indoor scenarios, as suggested by the U.S. Federal Communications Commission\cite{FCC23Usecase}.
Figure {\color{red}\ref{fig:landscape}} demonstrates these scenarios:
%

\begin{itemize}
    \item \textbf{Mobile broadband with licensed spectrum}. This wide-area mobile service is evolving from 5G to 5G-Advanced/6G for spectrum sharing.
    For instance, the SUs, i.e., mobile UE devices, collaboratively sense the upper mid-band Time Division Duplex (TDD) bands so that they can use the same frequency for uplink and downlink at different times. 
    Another example is that the downlink frequency used by base stations is the PU and the downlink frequency used by UE to UE communication is the SUs.
    The SUs sense available channels to improve communication throughput.
    
    \item \textbf{Wireless broadband with unlicensed spectrum}. These wireless local area networks support SU devices in dynamically accessing available channels from the Wi-Fi $7-8$ and beyond to enhance transmission throughput.
    For instance, Wi-Fi 7 allows devices to transmit and receive data simultaneously across multiple channels (MLO). 
    Thus, adapting ML-enabled spectrum sensing with the orthogonal frequency division multiplexing access (OFDMA) protocol enables devices to utilize various available unlicensed spectrums, enhancing communication performance. 
    This complementary solution can also reduce the additional traffic caused by existing CSMA/CA protocols.
    
    \item \textbf{Local licensing/indoor uses}: This scenario involves private entities/enterprises such as campuses, stadiums, and factories/warehouses, seeking to maximize the usage of spectrum availability in their geographic area given the mid-band ($7.1- 24$ GHz) and mmWave ($24-100$ GHz). It is also suitable for indoor use to minimize outdoor incumbent interference.
\end{itemize}

%

When the CSS is deployed in 6G networks, the attack surface is broader compared to the 5G-Advanced deployment in all spectrum sharing scenarios mentioned above. This is primarily due to the increasing number of base stations (PUs) with smaller coverage areas, as well as the growing number of SUs. 
Figure \ref{fig:cell_devices_trends} illustrates the trend of cells and authorized devices in Australia over the years. 
Similarly, in downtown Boston in the U.S., the 5G deployment is dense, with 50 gNBs covering a 1.5 × 1.5 km area for the upper mid-band spectrum ranging from $7-24$ GHz.
Table \ref{tab:attack_position} further breaks down FL and ML-based spectrum sensing studies concerning practical threat models in CSS design.

\noindent \textbf{Privacy-by-design at the FC server is not enough}.
Existing FL-enabled CSS studies often directly adapt federated aggregation to the FC to maintain the confidentiality of SUs' training data. 
These studies typically assume that the FC server is \textit{semi-honest} and controlled by a passive adversarial insider.
Intuitively, the attacker can access the local updates of SUs in every FL training round, whether it be model weights through Federated Averaging or the change in all weights (i.e., gradients) via Federated Stochastic Gradient Descent, without being able to manipulate the aggregated global ML model\cite{McMahan2016}. 
While this approach ensures confidentiality,   the privacy of SUs can still be exploited via the leakage of local gradient or weight update vectors.
This leakage is used to reconstruct the private training datasets of SUs, which can be achieved through optimization or analytics-based attacks.
For instance, gradient leakage attacks aim to minimize the distance between the shared gradients of SUs and dummy gradients by matching dummy training samples. 
On the other hand, analytic-based approaches recover the training inputs of fully connected layers in deep learning models by solving linear system equations.

Trusted Execution Environments (TEEs), e.g., Intel SGX or Arm TrustedZone, and Secure Multiparty Computation (MPC) are typically known as mitigation techniques to prevent gradient leakage. 
However, the massive deployment of these techniques to meet the demand of ultra-dense cellular networks may pose significant financial costs or efficiency and latency challenges. 
This is especially challenging as spectrum sensing is often considered a service in the Near-Real-Time RAN Intelligent Controller in 6G O-RAN (Open Radio Access Network). 
Nevertheless, compromising the FC is less practical than compromising the SUs, as the server is typically deployed in the control plane, requiring attackers to bypass other frontier protective components.

\begin{table}[!t]

\centering
  \caption{Recent 6G ML-enabled spectrum sensing studies} 
  \label{tab:attack_position}
\vspace{-4pt}

  \tabcolsep 0.02in
    \footnotesize{\halfcirc[2.5pt] \textemdash partially considered, \xmark \textemdash not considered}\\
    
    \begin{tabular}{c cc  cc  }
    
    \hline
     \multirow{3}*{Mechanism} &   \multirow{3}*{Scenarios} & \multicolumn{3}{c}{Threat Model}
 \\ 
    \cline{3-5}
 
 &    &  \textit{Compromised}  &  \textit{Compromised}  & \textit{External}  \\
  &      & FC Server &   SU/UEs   & Disruption  \\

   \specialrule{.1em}{.05em}{.05em} 
    FRDSA\cite{Dong24} & Mobile band &  \LEFTcircle & \xmark & \xmark \\
    FedSwap\cite{Gao21} & Mobile band &  \LEFTcircle & \xmark & \xmark \\
    FLAC\cite{Yang23} & Mobile band &  \LEFTcircle & \xmark & \xmark \\
    SecureFLCR\cite{Wasilewska23} & Wireless band  &  \LEFTcircle & \xmark & \xmark \\
    DeepSense\cite{Uvaydov21} & Wireless band&  \LEFTcircle & \xmark & \xmark \\
    DeepSweep\cite{robinson2024deepsweep} & Wireless band &  \LEFTcircle & \xmark & \xmark \\
    PartialObser\cite{Zhang24} & Local licenses &  \LEFTcircle & \xmark & \xmark \\
    
    \hline
  \end{tabular}

\vspace{-10pt}
\end{table}

\noindent \textbf{Enlarging attack surfaces by SU devices}.
%
With 6G expected to support connection densities ranging from one million to $100$ million devices per square kilometer, the number of SUs is projected to increase exponentially, as depicted in Figure \ref{fig:devices_trends}.
When adapting FL to the CSS systems, Byzantine poisoning attacks emerge as a typical threat model. 
In these attacks, data poisoning involves the attacker having access to the training data of compromised FL clients, while model poisoning requires access to the clients' models.
Typically, such poisoning attacks assume that the attacker gains control over a proportion of FL clients, i.e., SU devices.
This could be done either by Sybil attacks or additionally injecting fake clients or compromising existing ones.

Concerns arise from the fact that existing FL-enabled CSS studies, as presented in Table \ref{tab:attack_position}, have not addressed compromised SU devices. 
Through Byzantine poisoning attacks, such devices can collectively manipulate FL aggregation based on an adversarial objective function defined by the attacker. 
Consequently, the aggregated model at the FC becomes poisoned, exhibiting unexpected behavior on random testing labels (untargeted attacks) or focusing solely on the testing data aligned with the attacker's expected labels (targeted attacks).
Byzantine model poisoning attacks typically result in greater accuracy loss than data poisoning attacks.
However, they require the attacker to exert more adversarial engineering efforts to gain access to the local ML models of SUs.
Furthermore, theoretical FL studies have demonstrated that experimentally, a compromised client proportion of $20\%$ leads to a prediction accuracy loss of $30\%-80\%$ in the poisoned aggregated model. 
Nevertheless, in practice, these attacks are often negligible, given a small compromised proportion ranging from $0.1\%-1\%$ or less than $20\%$. 
Hence, poisoning the spectrum training data of as many possible SUs becomes a more practical threat concern in FL-enabled CSS systems.

\noindent \textbf{Disruptive technologies with jamming signals}.
Theoretical FL studies indicate that the prediction error rate of the poisoned aggregated FL model is directly proportional to the number of compromised FL clients, i.e., SUs.
Hence, poisoning the spectrum sensing of SU devices through jamming attacks emerges as a particularly promising approach for attackers. 
While adversarial jammers have been a typical concern in 4G and 5G/5G-Advanced deployments, the threat becomes even more significant in 6G deployment scenarios. 
This is primarily due to the capability of 6G transmitters to accommodate ultra-dense benign SUs in compact areas, potentially falling within the range of the attacker's jamming capabilities. 
For instance, existing jammers equipped with 8-14 antennas, capable of jamming Wi-Fi $5$GHz spectrum according to IEEE $802.11$ac/a/n standards, can generate a total output power of $18-40$W ($1-3$W per band) within a radius of $2-35$ meters. 
Additionally, compromised transmitter devices, particularly those that are full-duplex like industry-ready Universal Software Radio Peripheral (USRP) platforms, can also function as adversarial jammers.

\begin{table}[!t]

\centering
  \caption{Deep Learning Model Architecture} 
  \label{tab:model}
  \footnotesize{\normalsize FC: fully connected, C: convolution, P: Pooling, F: Flatten}
  \scalebox{1.}{
\normalsize
  \begin{tabular}{cccc}

\hline

\multirow{2}{*}{Architecture} & \multirow{2}{*}{Mechanism} & \multirow{2}{*}{Input Data} & Intermediate \\
& & & Layers\\
\hline
\multirow{2}{*}{DQN} & FRDSA\cite{Dong24} & State  & FC\\
\cline{2-4}
& FedSwap\cite{Gao21} &  State & FC \\
\hline
\multirow{2}{*}{FNN}  & \multirow{2}{*}{Ref. \cite{Shi22}} & Phase Shift & \multirow{2}{*}{Hidden}\\
& &  \&  RSS & \\
\hline
\multirow{3}{*}{CNN} & DeepSense\cite{Uvaydov21} & Raw I/Q & C, P, FC\\
\cline{2-4}
& \multirow{2}{*}{DeepSweep\cite{robinson2024deepsweep}} &  \multirow{2}{*}{Raw I/Q} &  C, P, Dense \\
\cline{4-4} & & & Drop, F\\
\hline
\end{tabular}
}
\end{table}

While recent studies on 6G spectrum sharing have received wide attention, poisoning spectrum sensing of SUs through jamming attacks has largely unexplored (see Table \ref{tab:attack_position}). 
Detecting jamming signals and protecting the integrity of the learning aggregation at the FC pose significant challenges. 
This can be attributed to two main factors. 
Firstly, adversarial jammers are not addressed in existing FL defense studies, which have traditionally focused on learning tasks such as computer vision and language processing. 
Consequently, the threat posed by compromised clients in these contexts is different from that of jammers in spectrum sensing systems. 
Secondly, jammers in spectrum sensing systems are inherently more difficult to detect compared to traditional denial-of-service (DoS) attacks and primary user emulation (PUE) attacks. 
This difficulty arises from the mobility of jammers equipped with portable batteries within the targeted wireless coverage area. 
Moreover, smart jammers can autonomously devise jamming attack strategies to evade periodic scanning detection approaches.

\section{Broaden ML Design for 6G Spectrum Sensing}

\noindent \textbf{Spectrum ML model specification.}
To realize the patterns of spectrum channel occupancy, FL-enabled studies typically deploy a deep learning model on each SU device wherein the model takes input as raw I/Q waveform data, radio signal strength (RSS), or channel states.
Then, the model's prediction output is the availability of multiple spectrum channels or a specific channel upon the model specification.
Table \ref{tab:model} presents widely used models in recent spectrum sensing studies including deep Q network (DQN), feedforward neural network (FNN), and Convolutional neural network (CNN).

Typically, the use of the DQN architecture enables SU devices to incorporate local reinforcement learning based on their input data states to derive spectrum scheduling plans.
This planning information is subsequently utilized to federatedly train the aggregated model at the FC server. 
For example, in FRDSA\cite{Dong24}, each SU collects metadata from historical transmissions to automatically guide local training, generating new model weights. 
Similarly, in FedSwap\cite{Gao21}, spectrum sensing data such as channel quality and request histories train the local model, which then outputs scheduling plans for the SU across multiple channels. 
In FNN implementations as used in\cite{Shi22}, each SU extracts phase shift and RSS from raw I/Q signals on a given channel, feeding them into the input layer of the local FNN model. 
Although all the above features can be extracted in $\mathcal{O}(1)$ computational time, recent approaches like DeepSense\cite{Uvaydov21} and DeepSweep\cite{robinson2024deepsweep} focus on raw I/Q signals to train a CNN model. 
This approach requires a consistent quadrature modulation setting among SUs and PUs.

\noindent \textbf{Real-time efficiency is 6G matter.}
Utilizing CNN for real-time spectrum hole detection receives more attention than previous DQN approaches.
The first reason is that CNN hardware accelerators can be adapted to SUs to accelerate local spectrum sensing at sub-millisecond speeds, utilizing unprocessed I/Q signals as inputs.
Secondly, the logits of CNN's output layer can be used to infer the availability of multiple bands simultaneously, similar to multi-label classification problems. 
This model design is compatible with SU and PU devices supporting multi-band Multiple Input Multiple Output (MIMO) antennas in 6G.

Experimentally, DeepSense\cite{Uvaydov21}  demonstrated that CNNs can accurately predict spectrum holes over $14$ channels of real-world IEEE $802.11$a transmissions and over $4$ channels with USRP N$210$ SDR devices. 
With LTE-M uplink transmissions simulating the Physical Uplink Shared Channel (PUSCH) over 16 bands, using a ($32\times2$) input tensor for I and Q signals as a sample, CNNs can predict spectrum holes with high accuracy.
Even with DeepSweep\cite{robinson2024deepsweep}, raw I/Q waveform signals are duplicated through a hardware I/Q mirror component to perform spectrum sensing without interrupting ongoing demodulation and decoding operations at both SU and PU devices.




%
%
%
%




\section{Conquer FL Integrity by Jamming Signals and Defence}

\noindent \textbf{Poisoning spectrum channels as attack targets}. 
Previous studies on federated learning (FL) integrity have demonstrated that Byzantine poisoning attacks can degrade prediction accuracy on either random labels (untargeted attacks) or specific labels (targeted attacks). %
This occurs because poisoning the local training datasets of SUs can manipulate the outcome of the aggregated FL model. 
Specifically, the FC server aggregates the local updates sent by both compromised and benign clients to generate a newly aggregated ML model for the next FL training iteration.

When mounting untargeted attacks on spectrum sensing applications, the goal is to make the aggregated model less reliable. 
This can result in either missing the detection of spectrum holes or incorrectly predicting the availability of occupied channels. 
Consequently, the communication throughput of SUs is reduced if they are advised to utilize such channels.

In contrast, targeted attacks intentionally report high activity on a specific channel of the PUs, allowing the attacker to exclusively secure the channel for adversarial purposes. 
Alternatively, by reporting low PU activity on targeted channels, the attacker can interfere with the communication of SUs suggested to use those channels.

\begin{figure}[!t]
\centering
\includegraphics[height=4.8cm]{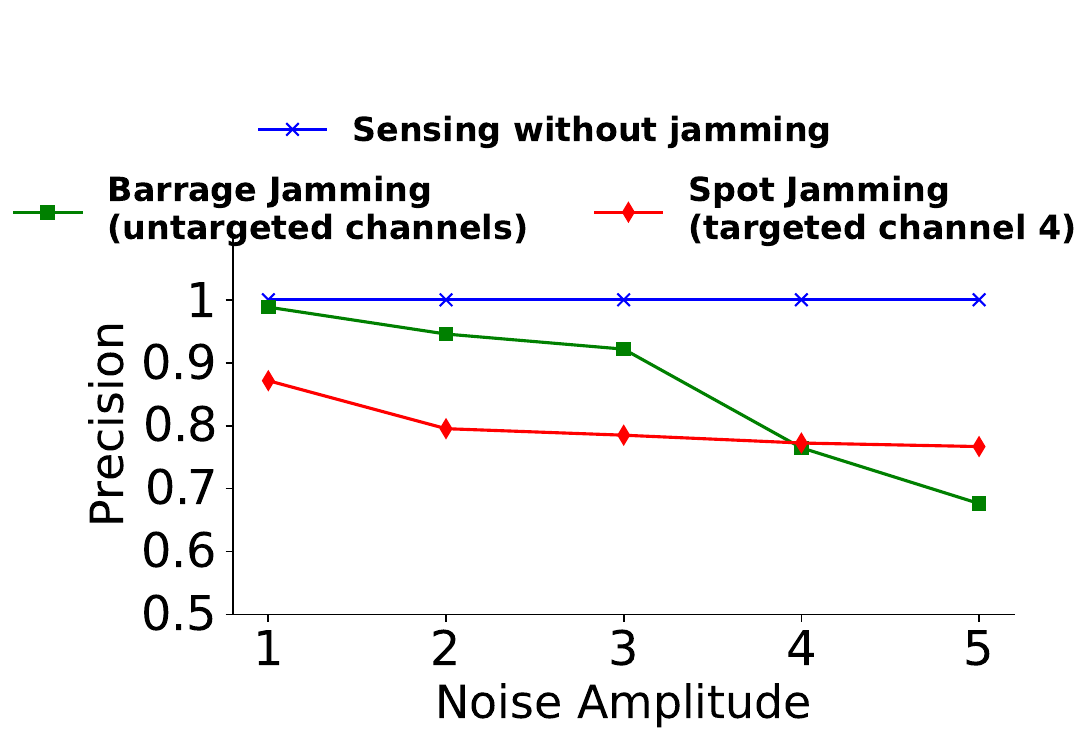}
\caption{Spectrum sensing precision degradation with LTE-M PUSCH uplink  over 4 channels using CNN model.}
\label{fig:jamming}
\end{figure} 

\noindent \textbf{Conventional jamming with untargeted/targeted channels.}
Individual SU devices acquire I/Q signals from nearby transmitter devices to train their corresponding local spectrum prediction models. 
Consequently, jammers with a large jamming range can disrupt any wireless link between two SUs in an ultra-dense network. 
Once jammed, the local ML models at the two end SUs cannot accurately sense the available spectrum channels in the network. 
Alternatively, jammers with limited range may be positioned near a targeted SU receiver. 
These jammers can introduce noise signals or block the original radio signal received by the SU. 
As a result, the I/Q samples used to train the local model of the SU are poisoned.

Two widely considered jamming modes in 5G studies are barrage and spot jamming. 
Barrage jamming spreads power over multiple channels or frequencies simultaneously, while spot jamming emits concentrated power directed toward a single channel or frequency. 
Consequently, barrage jamming is akin to launching untargeted attacks, whereas spot jamming is used for targeted attacks. 
Figure \ref{fig:jamming} illustrates the performance degradation caused by a barrage jammer adding noise signals across four channels (FL untargeted attacks) and a spot jammer targeting channel 4 (FL targeted attacks), respectively. 
In the simulation, LTE-uplink transmission is simulated with a PUSCH channel-specific configuration and QPSK modulation, where a CNN model with a  $(32\times 2)$ tensor input extracted from raw I/Q signals.



 


%




\noindent \textbf{AI-enabled adversarial jammers.}
The next generation development of jammers in 6G networks can leverage AI decision-making to efficiently schedule jamming signal emissions. 
For example, an adversarial binary classification ML model could aid in such scheduling\cite{Shi2018SpectrumDP}. 
Training data for the model includes the ratio of signal power to noise power (SNR) obtained by passively scanning the targeted wireless area. 
The training label is positive if the jammer detects either ACK packets from SUs on a targeted channel or scanned energy exceeding a threshold. 
Once trained, the model guides the jammer through the jamming schedule, enabling the creation of multiple adversarial models—one for each targeted spectrum channel.

In 6G, AI-enabled jammers pose a greater challenge for detection compared to conventional barrage and spot jamming. 
This difficulty arises because these advanced AI-enabled jammers do not concentrate energy at specific times. 
Even in scenarios with ultra-dense networks, such as wireless broadband, advanced jammers can train multiple models simultaneously, each targeting a specific SU for the attacker's benefit.

\noindent \textbf{AI-enabled jammer defence in 6G era.}
AI-enabled jammers present significant challenges to traditional equipment  tasked with detecting excessive power energy in a coverage area.
These jammers are mobile and can emit adaptive power at scheduled timestamps.
In addition, directly applying existing FL integrity defenses to the spectrum sharing context is not applicable. 
Firstly, jammers are not included in typical FL threat models, which only consider Byzantine nodes and the aggregated server.
Secondly, existing FL integrity defenses typically employ a centralized approach, where the aggregated server attempts to eliminate malicious updates from Byzantine nodes, whether they are model weight parameters or gradients of local training loss.
While this approach may eliminate spectrum sensing reports from SUs compromised by jammers, it may not accurately sense the currently occupied and available spectrum channels at such SUs' locations.
Therefore, a critical challenge to future 6G security studies is that:

\begin{center}
\textit{How can AI-powered jammers from nearby SUs be efficiently and effectively mitigated}?
\end{center}

While local defense in FL studies has received little attention, the spectrum sensing application in 6G has the potential to attract both network and machine learning security communities. 
Leveraging local defenses from individual or groups of nearby SUs could be effective in detecting mobile jammers. 
Additionally, leveraging the power of CNN hardware accelerators and semiconductor devices, such as Field Programmable Gate Arrays (FPGAs), is essential for enabling trustworthy spectrum sensing with low latency.
For example, every SU can leverage its local training dataset and the aggregated model sent by the FC server to verify whether its new training dataset have been poisoned by the nearby jammers.

%
%
%


\section{Privacy Threat by Secondary Users and Defence}

\noindent \textbf{Traditional ML-based privacy attack adaption}.
Although recent studies on FL-enabled spectrum sensing are widespread, they have not addressed compromised SUs. 
In FL, the FC server sends the latest aggregated ML model to SUs in each training round for their local training. 
Consequently, an attacker with access to this model can launch traditional ML privacy attacks, such as membership inference and property inference attacks. 
Membership inference attacks determine if a spectrum sensing sample was used to train the model, while property inference attacks infer the relationship between different attributes of the samples. %
These revealing attributes include radio signal strengths, geographic locations, and historical transmission states, among others.

\noindent \textbf{FL-based privacy attack adaption}.
In FL, compromised SUs can provide the attacker with multiple snapshots of the aggregated model over the training rounds. 
Consequently, the attacker can adaptively optimize the poisoning samples used in the local training of these SUs. 
These combined inference and poisoning attacks hold promise for enabling the attacker to extract more statistical information from the global model in subsequent FL training rounds. For instance:

\begin{itemize}
\item The statistical distribution of radio signal strengths from SUs allows attackers to derive device hardware specifications and popularly visited geographic locations. 
Such attacks are directly motivated by adversarial location-based recommendation services.

\item The attacker can differentiate between SUs that recently joined the FL-enabled spectrum sensing system in two consecutive FL training rounds. 
Consequently, the attacker can construct the trajectory based on the chronological order in which SUs are moving.

\end{itemize}

To conduct these privacy attacks, the attacker can model the correlations of gradient leakage (i.e., the loss after training samples) among compromised SUs across training rounds. 
This model can serve as a guide to craft new poisoning samples for the current training round. 
Even if the compromised SUs are not chosen in the current round, the attacker can utilize approximation optimization algorithms to estimate the training leakage.


\noindent \textbf{Privacy Defence with SUs.}
A natural approach to prevent gradient leakage during the local training of SUs is deploying TEEs, such as Trusted Armzone or Intel SGX, on devices. 
Unfortunately, this method is impractical for scenarios involving numerous SU devices in ultra-dense networks like wireless broadband. 
Alternatively, privacy-preserving machine learning solutions with distributed trust could be explored. For instance, SUs can interact with non-colluding edge servers to conduct privacy-preserving training. 
Another option is for individual SUs to fully decentralize their local training phase with other peers in a secret sharing setting. 
However, implementing this trust distribution requires initial cryptographic key management setups, and online computation time poses a challenge for real-time spectrum sensing in 6G. 
Therefore, a critical question worthy of investigation is:

\begin{center}
\textit{How can we effectively prevent gradient leakage and protect the confidentiality of FL aggregated models on local SUs?}
\end{center}

As 6G deployment is expected in the 2030s, designing new lightweight cryptographic schemes and deploying TEEs on SUs cannot wait.

\section{Conclusion}

This vision paper emphasizes practical security and privacy threats in typical 6G-enabled spectrum sensing scenarios. 
These threats encompass AI-powered adversarial jammers and compromised secondary users, presenting broader attack objectives compared to traditional 5G attacks. 
Defending against these threats poses even greater challenges than existing FL defense studies, which primarily concentrate on computer vision and natural language processing tasks. 
Therefore, we advocate that the development of the new 6G spectrum sharing system should prioritize security and privacy by design, rather than relying on ad hoc solutions.

%





%
%
%

\section{Acknowledgement}

This research paper is conducted under the 6G Security Research and Development Project, as led by the Commonwealth Scientific and Industrial Research Organisation (CSIRO) through funding appropriated by the Australian Government’s Department of Home Affairs. 
This paper does not reflect any Australian Government policy position. 
For more information regarding this Project, please refer to https://research.csiro.au/6gsecurity/.

\ifCLASSOPTIONcompsoc
  \section*{Acknowledgments}
\else
\fi


\bibliographystyle{unsrt}
\bibliography{fl}

\begin{thebibliography}{10}

\bibitem{Americas23}
Americas.
\newblock The evolution of 5g spectrum.
\newblock
  https://www.5gamericas.org/wp-content/uploads/2024/01/WP-Evolution-of-5G-Spectrum-1.pdf,
  2024.

\bibitem{NTIA24}
NTIA.
\newblock National spectrum strategy implementation plan.
\newblock
  https://www.ntia.gov/report/2024/national-spectrum-strategy-implementation-plan,
  2024.

\bibitem{Shi22}
Yi~Shi, Yalin~E. Sagduyu, and Tugba Erpek.
\newblock {Federated learning for distributed spectrum sensing in NextG
  communication networks}.
\newblock SPIE, 2022.

\bibitem{Wasilewska23}
Małgorzata Wasilewska, Hanna Bogucka, and H.~Vincent Poor.
\newblock Secure federated learning for cognitive radio sensing.
\newblock {\em IEEE Communications Magazine}, 2023.

\bibitem{McMahan2016}
H.~B. McMahan, Eider Moore, Daniel Ramage, Seth Hampson, and Blaise~Ag{\"u}era
  y~Arcas.
\newblock Communication-efficient learning of deep networks from decentralized
  data.
\newblock In {\em AISTATS}, 2016.

\bibitem{Minghong20}
Minghong Fang, Xiaoyu Cao, Jinyuan Jia, and Neil~Zhenqiang Gong.
\newblock Local model poisoning attacks to byzantine-robust federated learning.
\newblock In {\em {USENIX Security}'20}, 2020.

\bibitem{Chaudhari23}
H.~Chaudhari, J.~Abascal, A.~Oprea, M.~Jagielski, F.~Tramer, and J.~Ullman.
\newblock Snap: Efficient extraction of private properties with poisoning.
\newblock In {\em IEEE S\&P Oakland}, 2023.

\bibitem{FCC23Usecase}
Federal~Communications Commission.
\newblock A preliminary view of spectrum bands.
\newblock Online at
  \url{https://www.fcc.gov/sites/default/files/SpectrumSharingReportforTAC%20%28updated%29.pdf},
  2023.

\bibitem{Dong24}
Xuewen Dong, Zhichao You, Ximeng Liu, Yuanxiong Guo, Yulong Shen, and Yanmin
  Gong.
\newblock Federated and online dynamic spectrum access for mobile secondary
  users.
\newblock {\em IEEE Transactions on Wireless Communications}, 2024.

\bibitem{Gao21}
Zhihui Gao, Ang Li, Yunfan Gao, Bing Li, Yu~Wang, and Yiran Chen.
\newblock Fedswap: A federated learning based 5g decentralized dynamic spectrum
  access system.
\newblock In {\em 2021 IEEE/ACM ICCAD}, 2021.

\bibitem{Yang23}
Tongtong Yang, Wensheng Zhang, Yulian Bo, Jian Sun, and Cheng-Xiang Wang.
\newblock Dynamic spectrum sharing based on federated learning and multi-agent
  actor-critic reinforcement learning.
\newblock In {\em 2023 International Wireless Communications and Mobile
  Computing (IWCMC)}, 2023.

\bibitem{Uvaydov21}
Daniel Uvaydov, Salvatore D’Oro, Francesco Restuccia, and Tommaso Melodia.
\newblock Deepsense: Fast wideband spectrum sensing through real-time
  in-the-loop deep learning.
\newblock In {\em IEEE INFOCOM'21}, 2021.

\bibitem{robinson2024deepsweep}
Clifton~Paul Robinson, Daniel Uvaydov, Salvatore D'Oro, and Tommaso Melodia.
\newblock Deepsweep: Parallel and scalable spectrum sensing via convolutional
  neural networks.
\newblock In {\em IEEE ICMLCN 2024}, 2024.

\bibitem{Zhang24}
Weishan Zhang, Yue Wang, Xiang Chen, Lingjia Liu, and Zhi Tian.
\newblock Collaborative learning based spectrum sensing under partial
  observations.
\newblock {\em IEEE Transactions on Cognitive Communications and Networking},
  2024.

\bibitem{Shi2018SpectrumDP}
Yi~Shi, Tugba Erpek, Yalin~Evren Sagduyu, and Jason~H. Li.
\newblock Spectrum data poisoning with adversarial deep learning.
\newblock {\em MILCOM 2018 - 2018 IEEE Military Communications Conference
  (MILCOM)}, pages 407--412, 2018.

\end{thebibliography}

\end{document}